
\magnification=1200

\font \itp=cmti9

\font \bigbf=cmbx10 scaled\magstep2

\font \bigfonts=cmr9 scaled\magstep2
%
\hsize=15.0 truecm
\vsize=24.0 truecm
\baselineskip=7truemm plus 2truemm
%
%
%
\def\leaderfill{\leaders\hbox to 1em{\hss.\hss}\hfill}
\def\hb#1{\hbox to .6cm{\hss#1}}
\def\d{ \displaystyle}
\def\ltende{\longrightarrow}
\def\tende{\rightarrow}

\def\abs#1{\vert #1 \vert}

\def\la{\lambda}
\def\bye{\par\vfill\supereject\end}
\def\bl{\hfil\break}

\def\lb{\langle}
\def\rb{\rangle}

\def\df{d_{f}}

\def\om{\omega}

%
\def\abst#1{\centerline{
\hbox {\hsize=11.0 truecm
\baselineskip=4.5truemm minus 1truemm
\vbox to 1.0in{
\font \picc=cmr8
\noindent
\picc
 #1}}}\bl}
\def\abst1#1{\centerline{
\hbox {\hsize=13.5 truecm
\baselineskip=6.2truemm minus 1truemm
\vbox to 1.0in{
\font \piccc=cmr9
\noindent
\piccc
 #1}}}\bl}
\def\ref#1#2#3#4#5#6{\item{[#1]} #2, {\it #3 \/}{\bf #4}, #5 (#6).}
%
%
%
\ \ \hbox{\hskip 11.4truecm}DFBO-95-23\vskip 1.1truecm\noindent
\centerline{\bigbf Branching Interfaces with Infinitely Strong Couplings}
\vskip 1.7 truecm\noindent
\centerline{Giovanni Sartoni$^1$ and Attilio L. Stella$^2$}
\vskip 0.6truecm\noindent
\centerline{\itp
$^1$ Dipartimento di Fisica and Sezione INFN, Universit\`a di
Bologna, I-40126 Bologna Italy\/}\par\noindent
\centerline{\itp
$^2$ INFM--Dipartimento di Fisica and Sezione INFN,
Universit\`a di Padova, I-35131 Padova Italy\/}\bl\vskip 0.4truecm\noindent
\centerline{(July 27, 1995)}\vskip 1.4truecm\noindent
\centerline{\bigfonts Abstract}\vskip 0.9truecm\noindent
\abst1{A hierarchical froth model of the interface of a random $q$-state
Potts ferromagnet in $2D$ is studied by recursive methods.
 A fraction $p$ of the nearest neighbour bonds is made inaccessible to
domain walls by infinitely strong ferromagnetic couplings. Energetic and
geometric scaling properties of the interface are controlled by zero
temperature fixed distributions.
For $p<p_c$, the directed percolation threshold, the interface behaves as
for $p=0$, and scaling supports random Ising ($q=2$) critical behavior
for all $q$'s. At $ p=p_c$ three regimes are obtained for different ratios
of ferro vs. antiferromagnetic couplings. With rates above a threshold
value the interface is linear ( fractal dimension $d_f=1$) and its energy
fluctuations, $\Delta E$ scale with length as $\Delta E\propto
L^{\omega}$, with $\omega\simeq 0.48$.
When the threshold is reached the interface branches at all scales and is
 fractal ($d_f\simeq 1.046$) with $\omega_c \simeq 0.51$. Thus, at $p_c$,
dilution modifies both low temperature interfacial properties and critical
scaling.
Below threshold the interface becomes a probe of the backbone geometry
($\df\simeq{\bar d}\simeq 1.305$; $\bar d$ =
backbone fractal dimension ), which
even controls energy
fluctuations ($\omega\simeq d_f\simeq\bar d$). Numerical determinations of
directed
percolation exponents on diamond hierarchical lattice are also presented.
\vskip 6.2truemm\noindent
\hbox{Pacs numbers: 05.40.+j, 05.70.Jk, 75.10.Hk, 82.65.Dp}}
\vfill\supereject
{\bf 1 Introduction}\par\noindent
The effects of quenched disorder due to random impurities on phase
transitions are often non trivial and represent since long
an active research field.
According to a simple Harris criterion [1], disorder is relevant to
continuous phase
transitions with positive specific heat exponent. The presence of
impurities
can also change first order transitions into second order, and typically
has rounding
effects upon them [2,3]. Numerical and experimental evidence exists [4,5]
that in some random systems, like Potts ferromagnets,
the critical behaviour falls into the random bond Ising
model (RBIM) universality class, irrespective of the
different symmetry they possess.
Recently a justification of this possible universality has been
proposed, based on the analysis of the behaviour of critical interfaces in
presence of strong disorder [6].\bl
The interfacial free energy is indeed crucial to second order phase
transitions since it vanishes in a singular way at the critical point.
RBIM interfaces have been successfully
investigated within the solid-on-solid (SOS) model, which
ignores overhangs and islands near the interface [7].
This model is equivalent to the directed polymer in random medium
(DPRM) [8],
 which reduces to an optimization problem of path energies, at $T=0$.\bl
The $q$-state Potts model generalizes the Ising one with its $S_q$ symmetry.
When the model undergoes continuous phase transitions, Potts interfaces are
complicated
branched objects which become fractal at criticality, in the pure case. The so
called ``froth model'' has been introduced as a simplified description of those
interfaces [6]. This model represents a nontrivial generalization of the
DPRM and provides a possible mechanism for the explanation of the puzzling
universality in the strongly disordered Potts model at
criticality. The realization of the froth model on a diamond hierarchical
lattice
(DHL) in presence of coupling constant randomness is the only context where
the hard computational difficulties connected with its study could be
successfully faced so far. Indeed, on DHL the model is  amenable to an accurate
position space renormalization group analysis yielding deep insight
into interfacial scaling properties [6].\bl
A variant to the DPRM has been recently obtained through the introduction
of a geometrical
disorder represented by bond dilution [9], which shows to be itself an implicit
source of randomness for path energy distributions. Dilution, conceived as the
 presence of a fraction of bonds inaccessible by the polymer, modifies or
leaves unaffected DPRM
critical behaviour depending on whether the concentration of present
(accessible) bonds
is at or above the directed percolation threshold, respectively. The structure
of the underlying directed bond percolation cluster governs the scaling
behaviour of the model right at the critical threshold [9].\bl
Interest in investigating the effects of a similar dilution on the froth
interface model of ref.[6] is suggested by different considerations. As the
DPRM
can describe fracture [10], the froth interface can also be seen as realizing a
more complicated pattern of fracture, with the possibility of branchings
and loops formation in the cracks. Dilution can thus represent in both cases
the effects of hard inclusions through which cracks can not propagate, in an
inhomogeneous
material. In some systems like $^3$He-$^4$He mixtures in the pores of
aerogels, dense
coatings of $^4$He are formed on the surface of the aerogel structure [11].
The higher
 density considerably enhances the interaction between $^4$He atoms.
This effect adds to a more general introduction
of disorder in the interactions.
One expects that enhanced exchange interactions around the
aerogel prevent
penetration of the interface associated to the superfluid transition
in the aerogel [12].
The space regions occupied by the
aerogel and its coating
can again be seen as inclusions with very hard coupling for the
interface.\bl
These examples  suggested us to
consider a diluted generalization to the froth interfacial model of ref.[6],
by allowing
lattice bonds to be absent, i.e. unaccessible to the interface, with a finite
probability. As in ref.[9], we choose here a percolative dilution geometry,
which
is of course rather different from that of the aerogels' case.\bl
Dilution, especially at percolation threshold, is an obvious candidate
to possibly introduce
new universality classes of interfacial scaling behaviour. This is
most important for branching interfaces in random environment, because their
study is still in a very preliminary stage [6].\bl
A further motivation for diluting the froth interface model comes from the
legitimate hope that some of its scaling properties at percolation threshold
could be
more or less directly related to the geometry of the critical percolation
cluster.\bl
This article is organized as follows. In the next section we introduce the
froth model of the Potts interface and set up an iterative renormalization
group (RG) analysis for its study on hierarchical lattice. We discuss
in particular the specific features implied by the presence of dilution.
In the third section we present our results for several regimes of dilution.
The fourth section is devoted to a summary and to concluding remarks.\par
{\bf 2 Diluted model of the Potts interface}\par\noindent
The interface between two distinct ordered phases of a $2D$ random bond
$q$-state
Potts ferromagnet at very low temperature is a slightly fluctuating line.
Rising the
temperature to approach its critical value, bubbles of any of the other
$q-2$ states
may form at the interface; neglecting isolated islands and overhangs, as in SOS
models of Ising surfaces, we are left to consider an aggregate of bubbles, the
froth, each bounded by two SOS surfaces. This froth represents
the original interface. In a random bond Potts problem the interface crosses
 (breaks) bonds with random couplings, thus getting a random energy
contribution at each
step. In addition to random couplings, here we suppose there exist, with
probability $p$,  steps inaccessible to the interface (dilution).\bl
Even starting from the above simplified picture, the corresponding interfacial
model is too hard to analyse. By confining  the interface to the bonds
of the DHL shown in Figure 2.1, we obtain a simplified version of
the froth model, where a position space RG strategy
can be worked out. This model was already studied in ref.[6] in the undiluted
case ($p=0$), with only random exchange effects.\bl
The interface partition function, $Z_n$, on the diamond hierarchical lattice
at its $n-$th construction
stage (longitudinal length $L=2^n$) is calculated iteratively.
In the pure case (no randomness and no dilution),
$$Z_{n+1}= 2 Z_{n}^2 +(q-2) Z_{n}^4 \quad\eqno(2.1)$$
where $\d Z_0 =e^{-\beta 2J}$ is the Boltzmann weight of a broken
bond, and $\d
\beta=1/k_B T$. The entropic factor $q-2$ accounts for the number of different
coexisting phases which can occupy a bubble. An  analysis of the RG
flow described by eq.(2.1)
shows that there is a finite unstable fixed point $Z_c$, besides $Z=0$ and
$Z=\infty$. The quantity $\d f= \lim_{n\tende\infty}
\ln (Z_n) /L$  should be $<0$ and coincide with the interfacial line tension
for
$T<T_c$. So, the region $Z<Z_c$ corresponds to the low temperature regime,
because there
$Z= \exp \left( Lf\right)$, where $f$ ($f<0$) is the line
tension. For $Z>Z_c$ the line tension
description does not apply, since upon iterating eq.(2.1) $Z$ approaches
infinity
as $\exp \left( L^2f_b\right)$, where $f_b>0$ is a sort of dense froth
free energy which  describes the system in the high-T phase.
The RG analysis
provides the critical Boltzmann weight, $Z_c$,
and the interface critical exponent $\mu (q)$,
characterizing the vanishing of the interfacial free energy, $\d f\sim
{|T-T_c|}^{\mu}$, for any $q$ [13].
Indeed, if we
 put $t_n=Z_n-Z_c$, we get by construction $\d f(t_0)=2^{-1}f(t_1)=
2^{-1}f(2^{y_T}t_0)$,
which leads to the expected scaling of $f$ for $t\tende 0$, with
$\mu=\ln ({\partial t_1
\over\partial t_0}) /\ln2=y_T$. By keeping the lengths of the system
finite and equal to $L$ and $L'=L/2$, we also obtain the finite size
scaling (FSS) version of the previous law: $f(t_0,L)=2^{-1}f(2^{y_T}t_0,L')$.
Of course, the thermal exponent $y_T$ has
also the meaning of fractal dimension of the critical interface because
the average number of broken bonds is obtained by differentiating
$\ln Z_n$ with respect to $t_0$.\bl
In presence of random couplings and bond dilution, $Z_n$
is a random variable whose
distribution function, $\d P_n (Z)$, has to be iterated starting from the
initial form:
$$ P_0 (Z)=p \delta(Z)+(1-p){\cal P}_0(Z) \quad .\eqno(2.2)$$
Namely each accessible bond is given a weight, $Z$, according to ${\cal P}_0
(Z)$.
$p$ is the probability that a given bond is not accessible to the interface.
For a given
realization of disorder (2.2) in the whole structure, the partition function
evolves from a stage to the next one as:
$$Z_{n+1}=Z_n (1)Z_n (2)+Z_n (3)Z_n (4)+(q-2)\d\prod_{i=1}^{4}Z_n (i)
\quad ,\eqno(2.3)$$
where the indices from $1$ to $4$ refer to the four $n$-order elements
constituting an $n+1$-order diamond.
 $\d P_n (Z)$ iterates according to the rules of the nonlinear composition of
independent random
variables defined in (2.3). Fixed points of the mapping (2.1)  are replaced by
fixed distributions.\bl
It is easy to see that the component of $P_n$ concerning
the probability for a bond to be infinitely hard decouples from the other
quantities and iterates separately [9] as:
$$\d p_{n+1}=\left[1-(1-p_n )^2 \right]^2 \quad ,\eqno(2.4)$$
where $p_n$ is the probability for inaccessible macrobonds to
occur through the $n-$level
lattice. So (2.4) is nothing but the RG transformation for percolation
on DHL, whose critical
fixed point is at $\d p_c =\left(3-\sqrt{5}\right)/2$ [9].
As we will also discuss at the end of the next section, this is
a reasonably good qualitative model of directed percolation in $2D$.\bl
The second component, $\d{\cal P}_0(Z)$, of $P_0 (Z)$
is related to
energy randomness, and couples to bond dilution under RG flow.
Hence even in non random
energy cases, $\d {\cal P}_0 (Z)=\delta (Z-Z_0 )$, $Z_1$ no longer
possesses a binary
distribution with values $0$ and $Z$, already after the first iteration. Thus
$P_n (Z)$ always evolves towards a multivalued distribution
function, like in the random energy model.\bl
In our model we verified that both dilution and disorder are relevant
perturbations
of the pure system fixed point of eq.(2.1). Furthermore, similarly to what
happens in the undiluted case [6], the impure system behaviours turn out to be
always controlled by $T=0$ fixed point distribution functions. This is similar
to what happens in analogous problems [8,14], where randomness
is relevant and energy optimization
 criteria prevail with respect to entropic considerations. Dilution,
here, is a further
source of disorder. So it is no surprise that the dilute random
system is governed by zero temperature strong disorder fixed-point
distributions
for all dilution regimes. We examine properties of these distributions in
detail below by considering from the start $T=0$ RG recursions.\bl
We are then concerned with the iteration of the interfacial energy distribution
starting from one which introduces absent (i.e. inaccessible)
bonds with
probability $p$ and present (i.e. accessible) ones with  probability
$1-p$. The energies $e$ of the latter at the $n=0$ level
are chosen according to a probability density: $\d {\cal P}(e)=u\delta (e-1)+
(1-u)\delta (e+1)$. In the pure system we had $e=2J$ for all bonds.
By decreasing $u$ we expect to reach criticality conditions similar
to those found by rising $T$ at nonzero temperature [6].
On a given sample, the interface at level $n+1$ can cross either
of the two lattice branches made by two linked $n-$macrobonds,
or form a bubble through both. The choice among these three
possibilities is only dictated by the criterion of
energy minimization, provided the $n-$macrobonds are globally accessible.
 Monte Carlo iteration of the energy distribution density
is relatively simple and fast [15]. Starting with a large
sample ($\sim 2\cdot 10^5$) of
bond energies we could iterate it up to $n\sim 20$ without propagating too
large
 statistical uncertainties due to finite samplings of the distributions.\par
{\bf 3 Numerical and exact results on RG flows}\par\noindent
RG flows confirm the existence of a critical percolation
threshold for absent bonds at $p=p_c$. This is
indeed the threshold above which the interface can not anymore
cross the system, unless at the cost of an infinite positive energy, even
for finite system sizes.
The numerical value
$\d p_c =0.3820\pm 0.0008$ compares quite well with the
theoretical one. As in ref.[9]
we find the influence of dilution on the interfacial critical
system to be qualitatively
different depending whether $p<p_c$, or $p=p_c$.\bl
In the $p<p_c$ regime forbidden regions do not percolate and
only isolated finite clusters of inaccessible
bonds exist. Clusters are fractal on scales shorter than their linear size.
On large length scales they are homogeneously distributed
and should not exert any particular influence on the interface, except as an
additional source
of disorder. We observe in the interface two distinct phases separated by
a critical value, $\d u_c$, of the disorder parameter $u$ ($u/(1-u)$
gives the ratio of ferro- vs. antiferromagnetic accessible bonds).
$u_c$ of course depends on $p$ and on the fact that we were setting
to unity the starting absolute $e$ value.\bl
When $u>u_c$ the interface is linear, its energy distribution has mean value,
$\lb E(L) \rb$, which shifts as $\d \lb E(L)\rb\simeq \d \lb E_0 \rb L \gg 0$
($\d L =2^n$ is the length of the DHL at level $n$) and a width, $\Delta E(L)
$,
growing as $\d L^{\om}$, with $\om\approx 0.30$. The interface is thus a
directed path on a DHL with the energy distribution discussed in [16].
For $u<u_c$  we find a dense foam of bubbles, whose energy
density function has mean value
$\d \lb E(L) \rb\propto -L^2$ and width $\d\Delta E(L) \propto L$.
The central limit
theorem applies  to this distribution function which should
be asymptotically Gaussian.
At $u=u_c$  we have $\d\lb E(L) \rb\propto L^{y_c}$ and $\d\Delta E(L)
\propto
L^{\om_c}$, with $\d y_c \simeq\om _{c}\approx 0.32$, suggesting
that on approaching
$u_c$ the average interface energy should scale as:
$$\d\lb E(p,u,L)\rb =A\abs{u-u_c}^\mu L+B L^{\om_c} \quad .\eqno(3.1)$$
Our numerical estimate is $\mu=0.93\pm 0.08$.
By further considering that the critical interface is still linear,
since its mass,
$M(L)$, scales as $\d M(L)\propto L$,
we find here
remarkable coincidences with the undiluted random bond Potts interface [6],
suggesting
that the observed transition should fall in the same universality class.
Indeed, the linearity of the interface, and the value of $\mu$ compatible
with unity suggests that the randomness produces Ising critical behavior
independent of $q$ as discussed in ref.[6]. This conclusion is corroborated
by a study of the probability $P_l(n)$ that the interface forms a loop at
stage
$n$. Like in the undiluted case, this probability tends to $0$ as $n$
approaches infinity.\bl
In the critical percolating regime, $\d p=p_c $, both absent and present
bonds percolate and form incipient macroscopically
spanning clusters which have strong effects on the interface shape. Again
two phases exist, separated by a critical value, $u_c$, of $u$.\bl
The interface is linear in the region above $u_c$ and the energy
distribution function
flows under RG to a phase characterized by a
mean $\d \lb E(L) \rb\propto L$ and
a width $\d \Delta E(L) \propto L^{\om}$, with
$\d\om
=0.48\pm 0.02$, peculiar
of a directed path confined on an essentially one-dimensional
structure (the percolation
critical backbone).
This is precisely the regime found for the linear DPRM at the directed
percolation threshold in ref.[9].\bl
When $u<u_c$ a dense aggregate of bubbles forms, saturating
all paths which connect the ends of the DHL, hence invading the whole
percolation backbone
of present bonds. The energy has a distribution whose mean
and width scale as $\d
L^{y}$ and $\d L^{\om}$, respectively,
with $\d{y}=
{\om}=1.305\pm 0.006$.
Since the backbone has fractal dimension
$\d {\bar d}=1.30575......$ [17], we argue that the critical
percolation geometry strongly
controls the interface energy fluctuations. Indeed these are
of the same order as
those of the percolation cluster mass ($\d{\om}\simeq{\bar d}$).
The fact that the backbone mass and its fluctuations alone determine
the interface energy scaling properties is confirmed by the
results for $u=0$ (all bond energies equal and negative),
which yield the same $\d\omega\simeq{\bar d}$. Thus, backbone mass
fluctuations scale
like the average mass, and the randomness in bond energies
does not add new features to the interface energy fluctuations.\bl
At the transition point, $\d u=u_c =0.661\pm 0.001$, the scaling exponents of
$\d \lb E(L) \rb $ and $\d\Delta E(L) $ versus $L $ are, respectively:
$\d y_c
=0.50\pm 0.02$ and $\d \om_c=0.51\pm 0.01$. On approaching $u_c$ the average
interfacial energy fulfills a scaling of the form:
$$\d \lb E(p,u,L)\rb =A\abs{u-u_c}^{\mu} L+BL^{\om_c} \quad .\eqno(3.2)$$
with $\d \mu=0.94\pm 0.04$, as illustrated by the collapse of data
in Figure 3.1.
This value is not incompatible with the undiluted froth model
one [6]. However, collapse fits are not the only
way to estimate $\mu$.
In order to deeper investigate the nature of this
transition it is worthwhile
calculating also the scalings of interface mass, $M$, and probability of
forming loops ,
$\d P_l (n)$, at level $n$ [6]. $M$ grows as $\d {L}^{\df}$. $\df$
indicates the interface
fractal dimension, and
 we find in the three different regimes:
$$u>u_c \quad \df=1.000006\pm 3\cdot 10^{-6}\ \ ,$$
$$u=u_c \quad\ \ \  \df=1.046\pm 0.006\ \ \ ,$$
$$u<u_c \quad \df=1.3054\pm 6\cdot 10^{-4}\quad .$$
Notice that, as already mentioned, $\df\simeq{\om}\simeq{\bar
d}$ in the saturated phase, $u<u_c$ (see also Figure 3.2).\bl
The loop probability is asymptotically independent of $n$,
$\d P_l{\buildrel
{\scriptscriptstyle n\tende +\infty}\over \ltende} P_{\infty}$, and
$\d P_{\infty}=2^{\df -1}-1$ [6] provides an alternative way to
estimate $\df$.
The following behaviours are found under iteration ($L=2^n$):
$$u>u_c \quad ,\ \ P_l (n)\simeq A_1 e^{-bL} \hbox{ , }$$
which implies $ P_{\infty}=0$, hence $d_f =1$. Notice that
$A_1$ and $b$ depend of course on the actual $u$ value chosen.
%
$$u=u_c \quad ,\ \ P_l (n)\simeq P_{\infty}+{A_2 \over L^{\gamma}}
\hbox{ , }$$
with $\d\gamma =0.40\pm 0.01 $, $\d P_{\infty}=(2.3\pm 0.6)
\cdot 10^{-2}$, thus
$\d\df =1.03\pm 0.02$;
%
$$u<u_c \quad ,\ \ P_l (n)\simeq P_{\infty}-A_3 e^{-cL} \hbox{ , } $$
with $P_{\infty}=0.236\pm 0.0009$, which implies  $\d \df =1.306\pm 0.001$.
Unlike $P_{\infty}$, $A_3$ and $c$ show a dependence on $u$.\bl
Obviously there is a linear interface for $u>u_c$, while the interface
occupies the
whole percolation backbone when $u<u_c$. Indeed $\d \df\simeq{\bar d}$.
Most interesting is the fact that the interface
is no longer linear at the critical point $u=u_c$, because $\df$ is
significantly different from 1, in that case, and
the loop formation probability does not behave as a marginal field,
as it happened in the undiluted case [6].
Since $P_l$ is a relevant scaling field, logarithmic corrections existing
in the undiluted model disappear, and the critical interface
is a branched object, thus
indicating that we are in presence of a different, new type of
criticality.\bl
Moreover the RG mappings of $P_n (Z)$ at finite temperature evolves towards
$T=0$ distributions both within stable phase regions of parameter space,
and on
approaching the critical surface, showing that these phase transitions
 are also governed by zero temperature strong
disorder fixed points, as anticipated.\bl
For $p>p_c$ there are of course no paths joining the ends of the DHL,
because accessible bonds do not percolate anymore.\bl
%
%
Let us now discuss in more detail the percolative landscape across which
the interface develops when $p=p_c$.
The probability of a present bond is $\varrho=1-p$, which transforms
under RG as:
$$\d \varrho_{n+1}=\varrho_{n}^2 (2-\varrho_n )^2 \quad .\eqno(3.3)$$
The fixed point $\varrho_c =1-p_c$ sets the percolation threshold. So at
$p=p_c$ a critical cluster
of present bonds also exists, as already noticed. Considering the anisotropy of
DHL, which naturally introduces a preferred direction, the longitudinal one,
for
the percolation process (RG recursions (2.4) and (3.3) are explicit
consequences
of such a directedness), and taking into account
the DHL fractal dimension, equal to 2, the critical percolating cluster
on it is
eligible to represent a sufficiently faithful approximation of
the actual structure
of the $2D$ directed percolation cluster. The better such a circumstance
can be confirmed, of course, the more value should be added to our
investigation of interfacial properties on DHL.
Of course, there are limits to such an attitude: it should also not be
forgotten, for example,
that isotropic and directed percolation are not distinguishable on the
DHL.\bl
Since the Potts interface in its saturated phase
occupies the whole directed percolation backbone, we are able to use
such interface as a probe of backbone structure by the RG analysis.
The directed backbone fractal dimension, $\bar d$, has already been determined
both
analytically [17] and numerically by evaluating the saturated interface mass
scaling, shown in Figure 3.2.
We found a value surprisingly close to the one known for two-dimensional
directed percolation (${\bar d}\approx 1.31$ [18]).\bl
The correlation exponents of percolation,
$\d \nu_{\scriptscriptstyle \|}$ and $\d
\nu_{\scriptscriptstyle \bot}$, are equally accessible on DHL.
{}From (3.3) we calculate the rescaling of $\d \Delta \varrho=
\varrho -\varrho_c$:
$$\d \Delta \varrho'=\la_c \Delta
\varrho =b^{y_{\varrho}}\Delta \varrho \ \ ; \ \ \la_c =6-2\sqrt{5} \ \ .
\eqno(3.4)$$
The two correlation lengths, parallel and normal to the preferred direction,
diverge near $\varrho_c$ as:
$$\d \xi_{\scriptscriptstyle \|}\sim\abs{\varrho
-\varrho_c}^{-\nu_{\scriptscriptstyle \|}} \ \ \ ; \ \ \
\xi_{\scriptscriptstyle \bot}\sim
\abs{\varrho -\varrho_c}^{-\nu_{\scriptscriptstyle \bot}}\ .\eqno(3.5)$$
Manifestly, on this hierarchical lattice
$\xi_{\scriptscriptstyle \|}$ concerns the correlation
in the longitudinal direction connecting the two ends of the lattice.
Thus, if $\d \xi_{\scriptscriptstyle \|}' =b^{-1}
\xi_{\scriptscriptstyle \|}$, with $b=2$ (length rescaling factor), it follows:
$$\d {1\over \nu_{\scriptscriptstyle \|}}={\ln\la_c \over \ln b}=0.61151617...
\hbox{ \ \ and \ \ }\nu_{\scriptscriptstyle \|}=1.6352797...\ \ ,$$
(the two-dimensional exponent is slightly larger, $\d
\nu_{\scriptscriptstyle \|}=1.735$ [19]).\bl
To derive $\nu_{\scriptscriptstyle \bot}$ we define
the distance along the direction orthogonal
to the lattice, as equal to the number of branching levels separating two
given sites. With such a convention
the average transverse size of the percolation backbone,
$L_{\scriptscriptstyle \bot}$, transforms then according to:
$$L_{\scriptscriptstyle \bot}^{\scriptscriptstyle (n+1)}={2\varrho_{n}^2
(1-\varrho_{n})^2
 L_{\scriptscriptstyle \bot}^{\scriptscriptstyle (n)}
+2\varrho_{n}^3 (1-\varrho_{n})3L_{\scriptscriptstyle
\bot}^{\scriptscriptstyle (n)}+
\varrho_{n}^4 2L_{\scriptscriptstyle \bot}^{\scriptscriptstyle (n)}
\over \varrho_{n+1}}\ .\eqno(3.6)$$
Applying finite size scaling to (3.5) we get
$\d \xi_{\scriptscriptstyle \bot}\sim
L^{\nu_{\scriptscriptstyle \bot}/
\nu_{\scriptscriptstyle \|}}$, at $\varrho =\varrho_c$. From (3.6)
it follows $\d L_{\scriptscriptstyle \bot}^{\scriptscriptstyle (n+1)}
\Big\arrowvert_{\varrho_c} =4\varrho_{c}^{2}
L_{\scriptscriptstyle \bot}^{(n)}\Big\arrowvert_{\varrho_c}$,
then $\d {\nu_{\scriptscriptstyle \bot} \over
\nu_{\scriptscriptstyle \|}}={\ln 4\varrho_{c}^2 \over \ln 2}=
{1\over \nu_{\scriptscriptstyle \|}} $, \
and $\d \nu_{\scriptscriptstyle \bot}=1$ . These results should be
compared with the two-dimensional exponents:
$\d {\nu_{\scriptscriptstyle \bot} \over \nu_{\scriptscriptstyle \|}}=0.633$;\
\  $\d
\nu_{\scriptscriptstyle \bot}=1.097$ [19].\bl
Finally the simplicity of our RG also allows for iterating the number of steps
on which the saturated interface is linear, and which therefore connect
different blobs of the
directed backbone.
{}From a percolation point of view those are just the ``red bonds''
forming one dimensional
links of the directed backbone. Like in isotropic percolation [20] one can show
that the directed backbone red bond mass, $M_R$, at $\varrho_c$ has a scaling
form
$\d M_R \propto L^{D_{\scriptscriptstyle R}}$, with $\d D_{\scriptscriptstyle
R}=
{1\over \nu_{\scriptscriptstyle \|}}$, when the system has linear size $L$. We
measured a numerical value $\d D_{\scriptscriptstyle R}=0.6111\pm 0.0006$,
which fairly
agrees with the above theoretic prediction of $\d {1\over
\nu_{\scriptscriptstyle
\|}}$. Figure 3.3 shows a plot of $M_R$ versus $L$.\bl
The percolation critical exponents calculated above for our simplified model
of directed percolation are remarkably consistent and close to the truly
two-dimensional ones, that we hope the same could apply to the scaling
properties of the Potts interface we have determined here.\par
{\bf 4 Conclusions}\par\noindent
In this work we addressed the critical behavior of a Potts ferromagnet in
the presence of both exchange disorder and dilution. The scaling properties
of the interfacial free energy were considered and dilution acted in the
form of percolative regions not accessible to the interface, due to
extremely strong ferromagnetic couplings. By restricting interfacial
configurations
to those which can be hosted by a DHL, a very accurate RG analysis of the
line tension can be carried on. For all dilution regimes considered
interface scaling turns out to be controlled by $T=0$ fixed point
distributions. This makes the analysis relatively more easy, since $T=0$
recursions
for the interfacial energy can be studied more effectively than their
$T>0$ counterparts.\bl
When accessible regions are above their percolation
threshold ($p<p_c$), dilution does not seem to modify interfacial scaling
with respect to the undiluted case.
The three regimes occurring in this case are those on the basis of which
the Ising-like nature of criticality in disordered Potts ferromagnet
in $2D$ could be argued in ref.[6].
New scalings occur when dilution is at threshold and the interface has to
develop within the incipient infinite cluster backbone. When, at $T=0$,
$u$ is such to guarantee a sufficient dominance of positive energy bonds,
the interface remains linear at large scales and appears
to behave in the same way a strictly linear DPRM would in such an environment
[9]. Like in the corresponding regime above  percolation threshold,
ramification is fully inhibited
if $u>u_c$. However, the exponent $\omega\simeq 0.48$ falls in a new
universality
determined by the backbone geometry limiting the linear interface [9].\bl
A completely new behavior is realized right at $u=u_c$ when dilution
is at threshold.
In such conditions we get clear evidence that the interface has nonzero
probability of branching and thus behaves as a fractal. Its dimension slightly,
but definitely exceeds $1$.
Of course, by construction, on DHL a strictly linear interface can not behave
as a fractal. However, a fractal dimension equal to $1$
would persist in the linear regime
when replacing the DHL with more complicated lattices
allowing in principle for a fractal linear DPRM geometry.\bl
The above interface scaling regime at $u=u_c$ is the first critical one
with nontrivial fractal geometry met so far in this kind of studies, and
indicates the highly nontrivial effect that backbone geometry can exert
on the interface and on the critical behavior of the system.
So far, borderline regimes between the linear and maximally branched ones
in such froth models were never seen to allow for nonzero, intermediate
looping probability. Thus the critical behavior of the interface could not
be seen to be different from the linear, Ising-like one at low-T [6].
The values of $\mu$ and $d_f$ estimated for this regime lead to
$\nu=1/{\mu}\approx 1.06$ or $\nu=d_f\approx 1.046$ for a critical Potts model
subject to threshold dilution and disorder. This can be concluded within
the same limits of the arguments developed in ref.[6]. Like in that undiluted
case, the value of $\nu$, being associated to a $T=0$ fixed distribution,
is universal with respect to $q$.\bl
In the last regime, with critical dilution and $u< u_c$, we could realize how
the interface can become a sort of probe for the structural properties
of the backbone. The interface mass $M$ scales as the backbone mass
with longitudinal distance. Quite remarkably, the energy fluctuations are fully
determined by backbone mass fluctuations here. These backbone mass fluctuations
in turn scale with the same exponent as the total average mass.\bl
As an instructive exercise adding credibility to our DHL model
as a good qualitative picture of the situation on $2D$ Euclidean lattice,
we produced here results for $\nu_{\scriptscriptstyle \|}$
 and $\nu_{\scriptscriptstyle \bot}$ and the ``red bond'' dimension
of directed percolation. No similar determination of
$\nu_{\scriptscriptstyle \bot} $ was
produced before, to our knowledge. These results altogether appear
remarkably consistent with the numerical estimates in $2D$ [19].\par
{\bf Acknowledgement}\par\noindent
 One of us (A.L.S.) would like to thank Mehran Kardar and
 Nihat Berker for enlightening discussions.
\par\vfill\supereject
{\centerline{\bf References}}
\ref{1}{A.B. Harris}{J. Phys. C}{7}{1671}{1974}
\ref{2}{K. Hui and A.N. Berker}{Phys. Rev. Lett.}{62}{2507}{1989}
\ref{3}{A. Aizenman and J. Wehr}{Phys. Rev. Lett.}{62}{2503}{1989}
\ref{4}{S. Chen, A.M. Ferrenberg and D.P. Landau}{Phys. Rev.
Lett.}{69}{1213}{1992}
\ref{5}{L. Schwenger, K. Budde, C. Voges and H. Pfn\"ur}{Phys. Rev. Lett.}{73}
{296}{1994}
{\item{[6]} M. Kardar, A.L. Stella, G. Sartoni and B. Derrida,
{\it Phys. Rev. E (Rapid Comm.)}, in press (1995).}
{\item{[7]} G. Forgacs, R. Lipowsky and Th. Nieuwenhuizen, in
{\it Phase Transitions
and Critical Phenomena\/}, vol. 14, edited by C. Domb and J.L. Lebowitz
(Academic, New York, 1991).}
{\item{[8]} D.A. Huse and C.L. Henley, {\it Phys. Rev. Lett.\/} {\bf 54}, 2708
(1985); M. Kardar and D.R. Nelson, {\it Phys. Rev. Lett.\/} {\bf 55}, 1157
(1985); M. Kardar, {\it Phys. Rev. Lett.\/} {\bf 55}, 2923 (1985);
D.A. Huse, C.L. Henley and D.S. Fisher, {\it Phys. Rev. Lett.\/} {\bf
55}, 2924 (1985); M. Kardar and Y.-C. Zhang, {\it Phys. Rev. Lett.\/} {\bf 58},
2087 (1987).}
\ref{9}{L. Balents and M. Kardar}{J. Stat. Phys.}{67}{1}{1992}
{\item{[10]} J. Kert\'esz, {\it Physica \/}{\bf A191}, 208 (1992);
J. Kert\'esz,
V.K. Horv\'at and F. Weber, {\it Fractals\/} {\bf 1}, 67 (1993).}
\ref{11}{A. Falicov and A. N. Berker}{Phys. Rev. Lett.}{74}{426}{1995}
{\item{[12]} In a Bulme-Emery-Griffiths description (see ref.11) this
interface is like an Ising $+/-$ domain wall.}
{\item{[13]} As already stressed in ref.[6], this approach is not
able to show the
expected change to first order phase transition at $q=4$.}
{\item{[14]} M. Kardar, {\it Phys. Rev. Lett. \/}{\bf 55}, 2235 (1985).}
{\item{[15]} Starting form an ensemble consistent with ${\cal P}(e)$, at the
$(n\hskip-1.6pt+\hskip-1.6pt1)-$th iteration groups of four $n-$macrobonds
are sampled and the least path-energy is chosen to be the actual
value of that $(n\hskip-1.6pt+\hskip-1.6pt1)-$macrobond.}
\ref{16}{B. Derrida and R.B. Griffiths}{Europhys. Lett.}{8}{111}{1989}
{\item{[17]}} Since
$\d {\bar d}={\ln \left( n(1-p_c )/1\right) \over \ln (2/1)}$,
where $\d n(1-p_c )$ is the average number of present
bonds forming the critical
 backbone, R.G. gives $\d {\bar d}=1.30575....$.
{\item{[18]} S. Roux and Y.-C. Zhang, {\it preprint\/} (1995).}
\ref{19}{J.W. Essam, K. de'Bell, J. Adler and F.M. Bhatti}{Phys. Rev. B}{33}
{1982}{1986}
{\item{[20]} A. Coniglio, {\it Phys. Rev. Lett. \/}{\bf 46}, 250 (1981); A.
Coniglio, {\it J. Phys. A \/}{\bf 15}, 3829 (1982).}\par\vfill\supereject
\centerline{ FIGURES}\bl
\item{FIGURE 2.1.} a) The construction rule of the diamond
hierarchical lattice. At every
stage each bond is replaced by a 4-bond cell. b) The
DHL at its second stage.\bl
\vskip 1.5truecm\noindent
\item{FIGURE 3.1.} Data collapse of $\d \lb E(u,L,p=p_c )
\rb /L^{{\om}_c}$ against
$\d \abs{u-u_c}^{\mu} L^{1-{\om}_c}$. We measure
$\mu=0.94\pm 0.04$ and $\om_c =
0.51\pm 0.01$. Scales are logarithmic on both axes.
\vskip 1.5truecm\noindent
\item{FIGURE 3.2.} Plot of the interface mass, $M$, versus lattice
longitudinal
length, $L$, in the saturated phase, $u<u_c$, and at critical
dilution $p=p_c$.
$M$ equals the directed percolation backbone mass, which thus
exhibits the same
scaling against $L$.\bl
\vskip 1.5truecm\noindent
\item{FIGURE 3.3.} Plot of the ``red bond'' mass, $M_R$, in the
critical percolation
backbone, versus the lattice longitudinal length $L$.
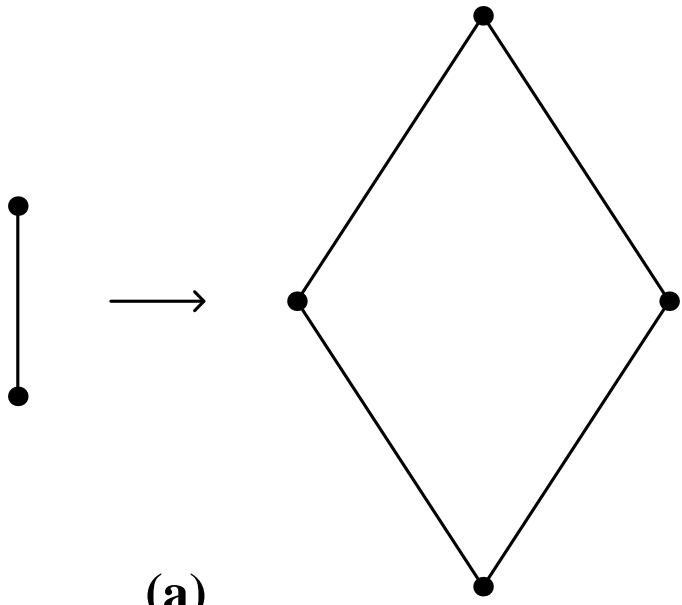
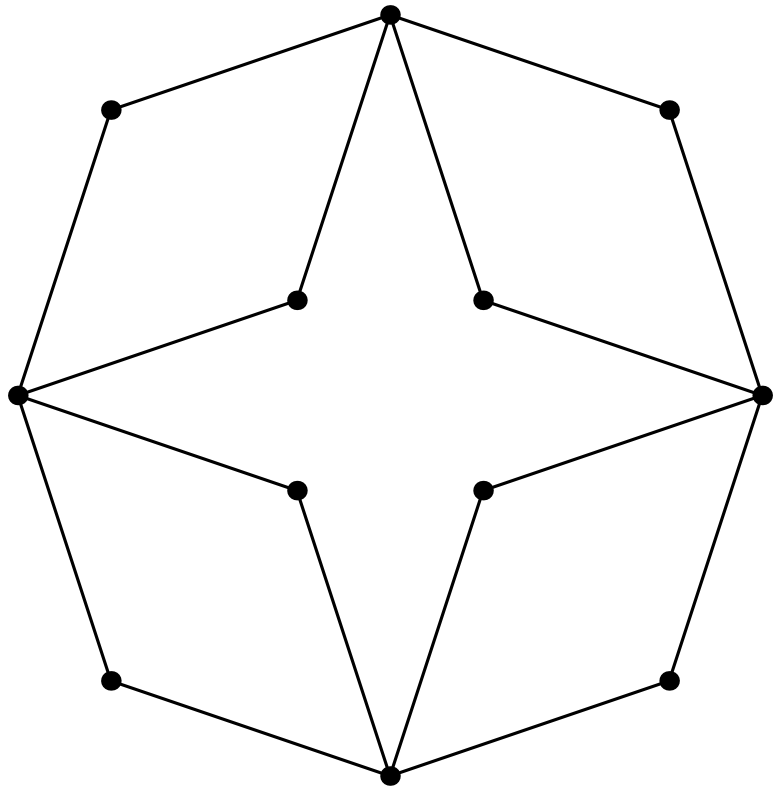
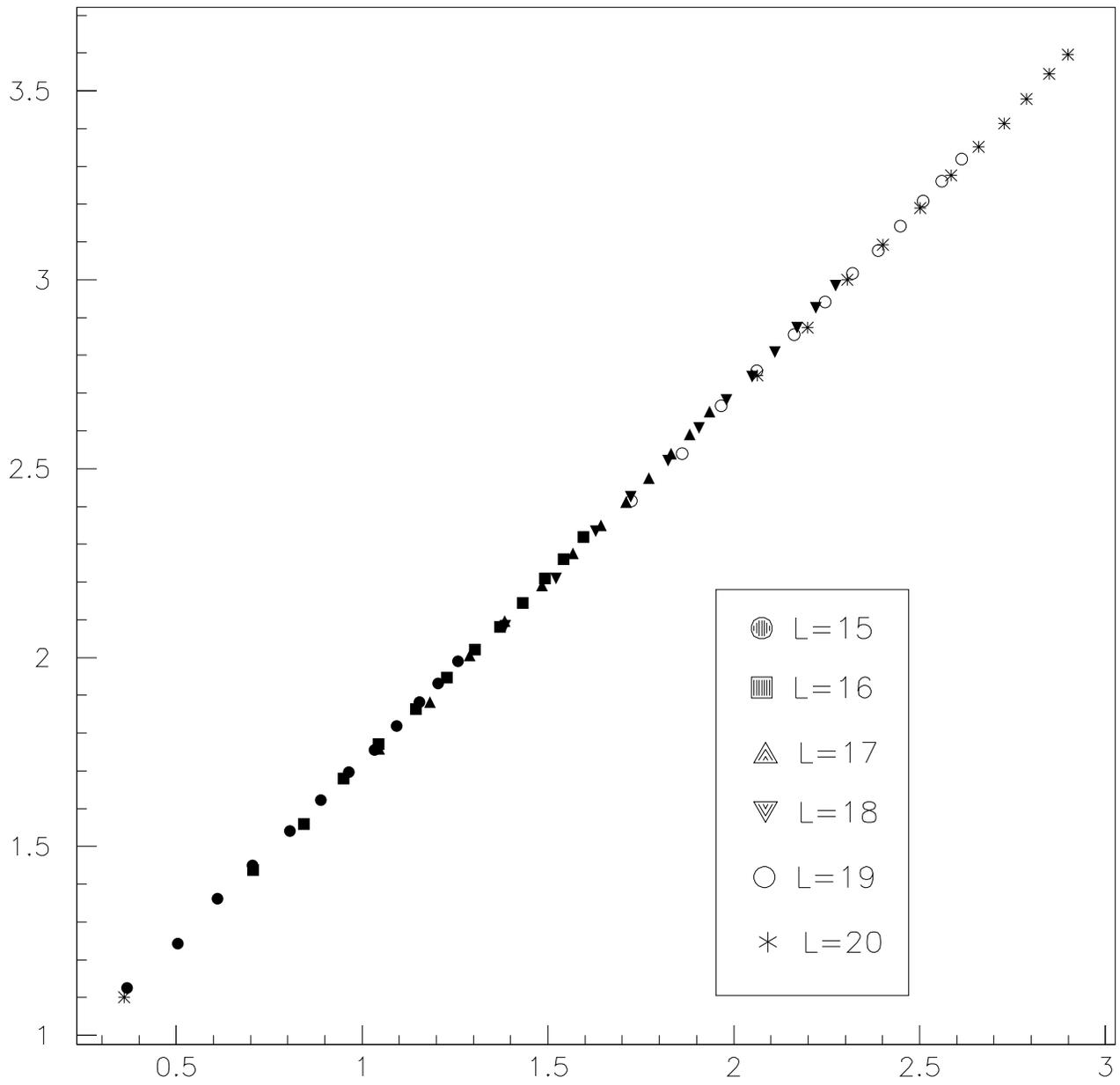
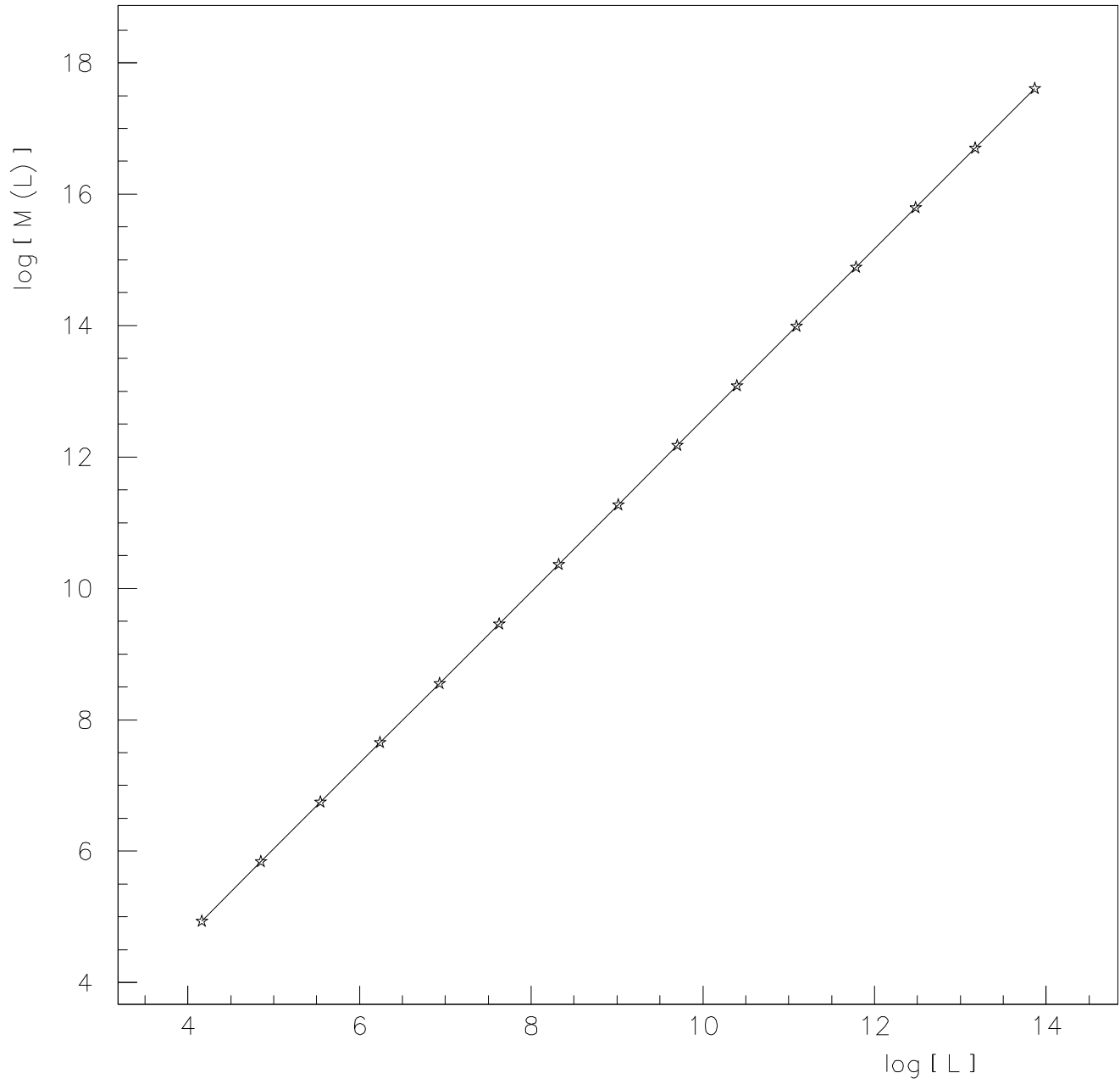
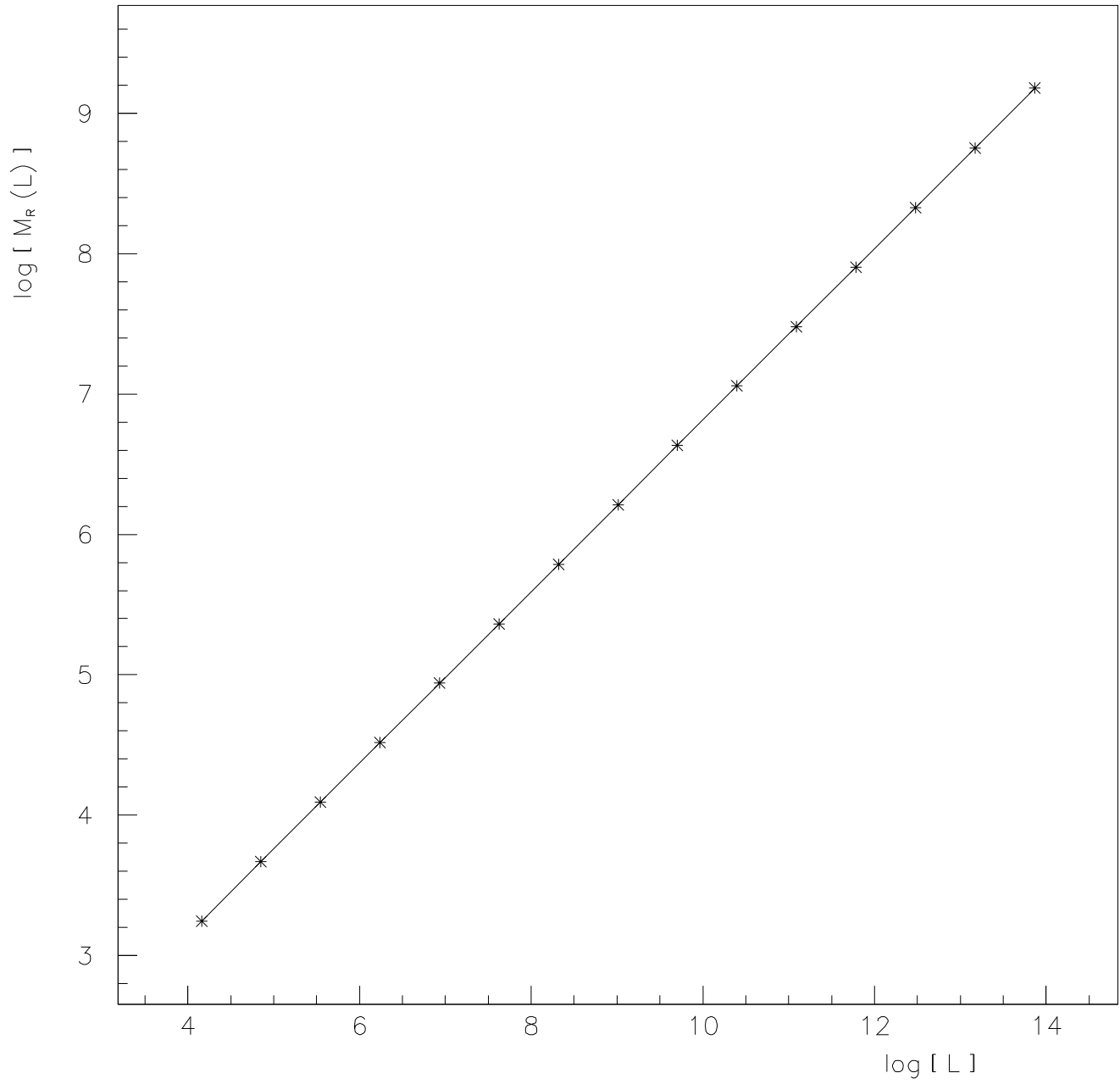
\bye